\documentclass[11pt,paper,a4paper]{JHEP3} % 10pt is ignored!

\JHEPspecialurl{http://jhep.sissa.it/JOURNAL/JHEP3.tar.gz}

\usepackage{epsfig,multicol,bbm}
\usepackage{amssymb}
\usepackage{amsmath}
\usepackage{cite}
\usepackage{epstopdf}
\newcommand\fverb{\setbox\pippobox=\hbox\bgroup\verb}
\newcommand\fverbdo{\egroup\medskip\noindent%
                        \fbox{\unhbox\pippobox}\ }
\newcommand\fverbit{\egroup\item[\fbox{\unhbox\pippobox}]}
\newbox\pippobox
\newcommand{\beq}{\begin{equation}}
\newcommand{\eeq}{\end{equation}}
\newcommand{\bea}{\begin{eqnarray}}
\newcommand{\eea}{\end{eqnarray}}
\newcommand{\bem}{\begin{multline}}
\newcommand{\eem}{\end{multline}}
\newcommand{\beg}{\begin{gather}}
\newcommand{\eeg}{\end{gather}}

\def\eq#1{{Eq.~(\ref{#1})}}

\def\eq#1{{Eq.~(\ref{#1})}}

\title{AAMQS: A non-linear QCD analysis of new HERA data at small-$x$ including heavy quarks}
\author{Javier L.  Albacete$^1$, N\'estor Armesto$^2$, Jos\'e
  Guilherme Milhano$^{3,4}$, Paloma Quiroga Arias$^{5}$ and Carlos A. Salgado$^2$
\vspace{0.1in}

{\it $^1$ Institut de Physique Th{\'e}orique, CEA/Saclay, 91191 Gif-sur-Yvette cedex, France.
URA 2306, unit\'e de recherche associ\'ee au CNRS.

 $^2$ Departamento de F\'{\i}sica de Part\'{\i}culas and  IGFAE,
Universidade de Santiago de Compostela, E-15706 Santiago de Compostela, Spain.

$^3$ CENTRA, Instituto Superior T\'ecnico (IST),
Av. Rovisco Pais, P-1049-001 Lisboa, Portugal.

$^4$ Physics Department, Theory Unit, CERN, CH-1211 Gen\`eve 23, Switzerland .

$^5$ LPTHE, UPMC Univ. Paris 6 and CNRS UMR7589, Paris, France.
 
\vskip 0.1in

E-mail addresses: {\tt javier.lopez-albacete@cea.fr, nestor.armesto@usc.es,
guilherme.milhano@ist.utl.pt, pquiroga@lpthe.jussieu.fr, carlos.salgado@usc.es}. } }
\preprint{CERN-PH-TH/2010-318}     % OR: \preprint{Aaaa/Mm/Yy\\Aaa-aa/Nnnnnn}

\abstract{ We present a global analysis of available data on inclusive structure functions and reduced cross sections measured in electron-proton scattering at small values of Bjorken-$x$, $x<0.01$, including the latest data from HERA on reduced cross sections. Our approach relies on the dipole formulation of DIS together with the use of the non-linear running coupling BK equation for the description of the small-$x$ dynamics. We improve our previous studies by  including the heavy quark (charm and beauty) contribution to the reduced cross sections, and  also by considering a variable flavor scheme for the running of the coupling. We obtain a good description of data, the fit parameters remaining stable with respect to our previous analyses where only light quarks were considered. 
The inclusion of the heavy quark contributions resulted in a good description of available experimental data for the charm component of the structure function and reduced cross section provide the initial transverse distribution of heavy quarks was allowed to differ (more specifically, to have a smaller radius) from that of the light flavors.}


\begin{document} 

%\setcounter{page}{1}
%%%%%%%%%%%%%%%%%%%%%%%%%%%%%%%%%%%%%%%%%%%%%%%%%%%%%%%%%%%%%%%%%%%%%%%%%%%%%
\section{Introduction} \label{intro}

The experimental data collected in electron-proton deep inelastic
scattering (DIS) experiments
%\cite{Adams:1996gu,Arneodo:1996qe},Abt:1993cb,Ahmed:1995fd,Aid:1996au,Adloff:1997mf,Adloff:1999ah,Adloff:2000qk,Derrick:1993fta,Derrick:1994sz,Derrick:1995ef, Derrick:1996hn,Breitweg:1997hz,Breitweg:1998dz,Breitweg:2000yn,Chekanov:2001qu,:2008tx}
at small values of Bjorken-$x$ constitute one of the most valuable
sources of information to test and explore the high-energy limit of
Quantum Chromodynamics (QCD).  It is a well settled theoretical result that in the limit of small Bjorken-$x$ or, equivalently, at high energies, deviations from standard collinear perturbation theory are expected on account of large gluon densities developing on the proton wave function. Such corrections can be interpreted in different ways. On the one hand, preserving unitarity of the theory sets an upper limit on the growth rate of the gluon densities in the proton. This limit is realized by the inclusion of gluon recombination processes, highly probable in a high density environment, into the high-energy evolution equations. Such task is best carried out in the framework of the Color Glass Condensate (CGC), which is equipped with a set of renormalization group equations, the BK-JIMWLK equations  \cite{Jalilian-Marian:1997gr, Jalilian-Marian:1997dw, Kovner:2000pt, Weigert:2000gi, Iancu:2000hn,Ferreiro:2001qy,Balitsky:1996ub,Kovchegov:1999yj}, that include the needed unitarity corrections. 
On the other hand, the interplay between radiation (linear) and recombination (non-linear) process gives rise to a  dynamical transverse momentum scale, the saturation scale, which signals the onset of non-linear corrections.  The presence of a perturbatively large transverse momentum scale would invalidate the main assumption of collinear perturbation theory, where partons are assumed to move collinearly with the hadron, with zero transverse momentum.

On the phenomenological side, intense activity has been carried out in order to identify the onset of non-linear corrections in available data. The dipole model formulation of deep inelastic scattering \cite{Nikolaev:1990ja,Mueller:1989st} is an instrumental tool in this search, since it allows for a relatively simple implementation of saturation effects in the description of the scattering process. Indeed, dipole models  \cite{Golec-Biernat:1998js,Iancu:2003ge,Gotsman:2002yy,Albacete:2005ef,Kowalski:2003hm,Kowalski:2006hc,Goncalves:2006yt} successfully describe a number of features observed in data at small-$x$.
%, from the small-$x$ behavior of inclusive and diffractive structure functions, including the feature of {\it geometric scaling} \cite{Stasto:2000er}, to the ratio of diffractive over total cross-section or exclusive processes like vector meson production. 
%yielding strong support to the idea that saturation effects are indeed present in available data. Despite their success, most of the dipole models are somewhat unsatisfactory from a theoretical point of view: Rather than calculated from first principles, the small-$x$ dynamics is modeled through the introduction of a $x$-dependent saturation scale, $Q_s^2(x)\propto x^{-\lambda}$, with $\lambda$ a free parameter fitted to data. 
Important progress has been achieved in reducing the degree of modeling required in phenomenological works thanks to the determination of higher order corrections to the BK-JIMWLK equations \cite{Balitsky:2006wa, Kovchegov:2006vj,Gardi:2006rp,Balitsky:2008zza}. The latter provide the most complete theoretical tool so far to describe the small-$x$ dynamics of the dipole scattering amplitude including unitarity corrections. While higher order corrections present a complicated structure and are not amenable to a numerical implementation, it was demonstrated in \cite{Albacete:2007yr,Albacete:2007sm} that considering {\it only} running coupling corrections to the BK equation, henceforth referred to as rcBK, grasps most of the higher order effects. It was also observed in these works that the quantitative features of the solutions of the rcBK equation, and the much milder $x$-dependence of the saturation scale \cite{Albacete:2005ef}, are compatible with the values extracted from purely phenomenological analysis, rising hope that the rcBK equation could be used as a phenomenological tool. This idea was confirmed in a previous work by some of the authors \cite{Albacete:2009ps, Albacete:2009fh}, where we demonstrated the ability of the rcBK equation to correctly describe the data available at the time on inclusive and longitudinal structure functions measured in e+p collisions. Later on, a similarly successful description data for the diffractive structure functions was presented in \cite{Betemps:2009ie}. The rcBK equation has also been successfully employed in the analysis of data on total multiplicities in Au+Au and LHC Pb+Pb collisions \cite{Albacete:2007sm, Albacete:2010ad} and on single \cite{Albacete:2010bs} and double inclusive \cite{Albacete:2010pg} spectrum in p+p and 
d+Au collision performed at RHIC, where saturation effects are believed to be an important dynamical ingredient. Thus, the rcBK equation which, we recall, follows from a strict derivation in pQCD in a given limit (see section 2 for details), has become the most effective phenomenological tool to assess the role of unitarity effects in available data in a theoretically controlled way, thus bridging the gap between theory and experiment.

On the other end of the theory spectrum, DGLAP based analyses (see
\cite{Dittmar:2009ii} and references therein)  have consistently reported good fits to e+p data for $Q^2\gtrsim1\div 4$ GeV$^2$. A relevant question is whether the flexibility in the initial conditions for DGLAP evolution is hiding some interesting QCD dynamics, namely the presence of non-linear behaviour. A recent work \cite{Caola:2009iy} from members of the NNPDF collaboration showed that there is tension in NLO DGLAP fits to HERA data on structure functions when data sets with $Q^2$ below the estimated saturation scale of the proton were excluded from the analysis. Moreover, it was found that such deviations between theory and data can not be corrected by the inclusion of next-to-next-to leading order corrections in DGLAP evolution nor by a better treatment of heavy quark effects. It has been suggested that this problem may be fixed by equipping DGLAP evolution with additional perturbative BFKL-like resummation of small-$x$ effects \cite{Altarelli:1999vw,Ciafaloni:2003rd}, which seem to work in the right direction to reconcile data and theory. However, no analysis of data using small-$x$ resummed DGLAP evolution has been performed to date. In any case, it should also be taken into account that DGLAP has no predictive power towards small-$x$, since all the $x$-dependence of the parton distribution functions is encoded in the initial conditions for the evolution. On the other hand, analyses relying on the use of $k_t$-factorization using an unified DGLAP/BFKL resummation approach --thus beyond collinear factorization but still in the linear regime--  also provide a good description of HERA data for the longitudinal structure function, which is more sensitive to saturation effects. Finally, models \cite{Capella:2000pe,Armesto:2010ee} based on the non-pertubative Regge calculus have also proven able to describe e+p data on structure functions.

Thus, and despite the indications for the possible presence of saturation effects outlined above, it is not clear from present analyses whether current data signals the breakdown of collinear factorization nor the onset of non-linear corrections to the QCD evolution equations. Further, it remains unclear what is the Êprecise kinematic region where these two regimes of QCD should be distinguishable. The answer to these questions demands either that new data on different observables at smaller values of $x$ is obtained --that is the purpose of the proposed experimental facilities as the LHeC \cite{lhec2} or the EIC \cite{EICwhite}-- or that the more accurate data obtained by H1 and ZEUS collaborations is analyzed

%The answer to these questions demands either that new data on different observables at smaller values of $x$ is obtained, and that is the purpose of the proposed experimental facilities as the LHeC \cite{lhec2} or the EIC \cite{EICwhite}, or that the more accurate data obtained by H1 and ZEUS collaborations is analyzed.

Such is precisely the goal of this work. Here we extend our previous analysis by including the recent data on reduced cross section in e+p collisions at HERA as given by the combined analysis of the H1 and ZEUS collaborations \cite{:2009wt}, which supersedes older HERA data on inclusive and longitudinal structure functions. Data from E665 and NMC experiments are kept in the analysis. Besides of extending the explored kinematic range, the recent measurements performed at HERA have two clear advantages over previous data sets. On one hand, the combination method employed in the analysis reduces the systematic uncertainties, resulting in improved accuracy and smaller error bars, thus posing more constraining conditions on models. On the other, the fact that data is given directly in terms of the reduced cross sections $\sigma_r$, which is the experimentally measured quantity, eliminates the theoretical bias in the extraction of $F_L$ and $F_2$ from data. 

Besides the analysis of new HERA data, the main novelty in this work is the inclusion of dynamical heavy quarks, charm and beauty. It is known from direct experimental measurements \cite{:2009ut,Adloff:1996xq,Adloff:2001zj,Aktas:2005iw,Aktas:2004az,Breitweg:1997mj} that charm and beauty contribute substantially  to the total $\gamma^*p$ cross section and, therefore, their contribution should not be neglected. Technically, the charm and beauty contribution can be obtained by simply extending the sum over quark flavors in the dipole model \eq{dm1}. However, as already noticed in \cite{Golec-Biernat:1998js,Soyez:2007kg}, such straightforward inclusion of heavy quarks has strong effects on the fit parameters. In particular it tends to strongly reduce the saturation scale of the proton. This is related to the fact that the large charm (or beauty) mass acts as an infrared regulator for the fluctuation of the virtual photon into a quark-antiquark dipole, thus leaving less room for interaction in the kinematic region where unitarity effects are expected to be important. Other fit parameters also change noticeably after such direct inclusion of heavy quarks, thus blurring their physical  interpretation. We notice that a simple modification on the assumptions concerning the normalization of the heavy quark contribution or, equivalently, concerning the average radius of their transverse distribution, fixes the problem of stability of the fits.  For a consistent treatment of dynamical heavy quark effects we consider not only its contribution to the DIS cross section but we also implement a variable flavor scheme for the beta function to properly incorporate such effects in the running of the coupling. 

This work is structured as follows: Section 2 is devoted to a brief review of our theoretical setup, which relies on the dipole model formulation of the e+p scattering process and in the use of the rcBK equation to describe the small-$x$ dynamics of the dipole scattering amplitude. There we discuss the free parameters in the fit, together with our choice of initial conditions for the solution of the rcBK equation. The implementation of the variable flavor number scheme for the running of the coupling as well as the infrared regularization of the coupling are discussed in section 2.2. The experimental data included  in the fits and the numerical method devised to perform the global fits are discussed in section 3.
Our results are presented in section 4, where we first present the fits including only light quarks in the analysis. We then include the effects of charm and beauty, finding in both cases a good description of data. 
Finally, we wrap up with summary and conclusions.

\section{Setup} \label{setup}
In this section we briefly review the main ingredients needed for the
calculation of the inclusive and longitudinal DIS structure functions, which was extensively discussed in our previous paper \cite{Albacete:2009fh}. Neglecting the contribution from $Z$ boson exchange, only relevant at $Q^2$ much larger than those considered in this work, the reduced cross section can be expressed in terms of the inclusive, $F_2$, and longitudinal, $F_L$, structure functions:
\begin{equation}
\sigma_{r}(y,x,Q^2)=F_2(x,Q^2)-\frac{y^2}{1+(1-y)^2}F_L(x,Q^2),
\label{rcs}
\end{equation}
where $y=Q^2/(s\,x)$ is the inelasticity variable and $\sqrt{s}$  the center of mass collision energy.
In turn, at $x\ll 1$, the inclusive and longitudinal structure functions can be expressed as
\begin{eqnarray}
F_2(x,Q^2)=\frac{Q^2}{4\,\pi^2\alpha_{em}}\left(\sigma_T+\sigma_L\right)\,,\\
F_L(x,Q^2)=\frac{Q^2}{4\,\pi^2\alpha_{em}}\,\sigma_L\,.
\label{f2l}
\end{eqnarray}
Here $\sigma_{T,L}$
stands for the virtual photon-proton cross section for transverse ($T$)
and longitudinal ($L$) polarization of the virtual photon. In the dipole model, valid at high energies or small $x$, one writes  \cite{Nikolaev:1990ja,Mueller:1989st}:
\begin{equation}
  \sigma_{T,L}(x,Q^2)=2\sum_f\int_0^1 dz\int d{\bf b} \,d{\bf r}\,\vert
  \Psi_{T,L}^f(e_f,m_f,z,Q^2,{\bf r})\vert^2\,
  {\cal N}({\bf b},{\bf r},x)\,,
\label{dm1}
\end{equation}
where $\Psi_{T,L}^f$ is the light-cone wave function
for a virtual photon to fluctuate into a quark-antiquark dipole of quark flavor $f$. Note that $\Psi_{T,L}^f$ only depends on the quark flavor $f$ through the quark mass $m_f$, and electric charge $e_f$ (see e.g.  \cite{Golec-Biernat:1998js} for explicit expressions to lowest
order in $\alpha_{em}$).  ${\cal N}({\bf b},{\bf r},x)$ is the imaginary part of the dipole-target scattering amplitude, with $\bf r$ the transverse dipole size and $\bf b$ the impact parameter of the collision.
The study of impact parameter dependence of the dipole amplitude is controlled by long-range, non-perturvative phenomena rooted in the physics of confinement and thus is not tractable perturbatively. Following other works, we shall assume an average over impact parameter through the replacement:
\begin{equation}
2\int d{\bf b}\rightarrow \sigma_0\,,
\end{equation}
where $\sigma_0$ has the meaning of (half) the average transverse area of the quark distribution in the transverse plane and will be one of the free parameters in the fit. Attempts to go beyond the translational invariant approximation in the BK equation have been recently presented in \cite{Berger:2010sh,Kormilitzin:2010at}.
Finally, we shall also include in the fits available data for the charm contribution to the inclusive structure function $F_{2c}$, which can be calculated by considering only the charm contribution in Eqs (2.2) and (\ref{dm1}). Details about the normalization and initial conditions for $F_{2c}$ are given below. In order to approach safely the photoproduction region, we shall also consider the standard kinematic shift in the definition of Bjorken-$x$ \cite{Golec-Biernat:1998js}: 
\beq
\tilde{x}=x\left(1+\frac{4\,m_f^2}{Q^2}\right).
\label{xtilde}
\eeq
The mass of the three light quarks is taken to be $m_{l}=0.14$ GeV in some cases or left as a free fit parameter,  whereas that of charm and beauty are taken to be $m_{charm}=1.27$ GeV and $m_{beauty}=4.2$ GeV respectively \cite{Nakamura:2010zzi}.

\subsection{BK equation with running coupling}
\label{bk}

The main dynamical input in this work is the rcBK equation, which corresponds to the large-$N_c$ limit of the full B-JIMWLK equations. It resums to all orders leading radiative corrections in $\alpha_s\,\ln(1/x)$ and also a subset of the full next-to-leading order corrections \cite{Balitsky:2008zza}, namely running coupling corrections. The impact parameter independent BK equation reads
%The CGC is equipped with a set of renormalization group equations, the BK-JIMWLK equations, which allow to describe the small-$x$ evolution of the dipole amplitude, and, apart from trivial kinematic factors, that of the reduced cross section and of the structure functions in  \eq{rcs} as well. The leading order BK equation \cite{Balitsky:1996ub,Kovchegov:1999yj}, which corresponds to the large-$N_c$ limit of the JIMWLK equation,  resums radiative corrections in $\alpha_s\,\ln(1/x)$ to all orders and also contains non-linear corrections ensuring unitarity of the theory. Only recently the next-to-leading order corrections to the BK equation have become available. They are, however, of a complicated structure and not amenable for numerical implementation. However, as argued in \cite{Albacete:2007yr} and demonstrated in our previous analysis \cite{Albacete:2009ps}, considering only a subset of the higher order effects, namely only running coupling corrections, renders the BK equation compatible with experimental data while keeping the relative simplicity of LO equation, since their inclussion can be achieved by just modifying the evolution kernel. The impact parameter independent BK equation reads
\begin{eqnarray}
  \frac{\partial{\cal{N}}(r,x)}{\partial\,\ln(x_0/x)}&=& \int d{\bf r_1}\,
  K^{{\rm run}}({\bf r},{\bf r_1},{\bf r_2})\nonumber \\
 &\times&
 \left[{\cal N}(r_1,x)+{\cal N}(r_2,x)-{\cal N}(r,x)-
    {\cal N}(r_1,x)\,{\cal N}(r_2,x)\right]\,,
\label{bkrun}
\end{eqnarray}
with the evolution kernel including running coupling corrections given by
\cite{Balitsky:2006wa}
\begin{equation}
  K^{{\rm run}}({\bf r},{\bf r_1},{\bf r_2})=\frac{N_c\,\alpha_s(r^2)}{2\pi^2}
  \left[\frac{r^2}{r_1^2\,r_2^2}+
    \frac{1}{r_1^2}\left(\frac{\alpha_s(r_1^2)}{\alpha_s(r_2^2)}-1\right)+
    \frac{1}{r_2^2}\left(\frac{\alpha_s(r_2^2)}{\alpha_s(r_1^2)}-1\right)
  \right]\,,
\label{kbal}
\end{equation}
where ${\bf r_2}={\bf r}-{\bf r_1}$ and $x_0$ is the value of $x$ where the evolution
starts. In our case $x_0=0.01$ will be the highest experimental value of $x$ included in the fit.

%%%%%%%%%%%%%%%%%%%%%%%%%%%%%%%%%%%%%%%%%%%%%%%%%%%%%%%%%%%%%%%%%%%%%%%%%%%%%%%
\subsection{Variable flavor scheme and regularization of the coupling}
\label{ir}

The coupling in the rcBK kernel  Eq. (\ref{kbal}) is given, for a given number of active quark flavors $n_f$, by 
\begin{equation}
\label{eq:rc}
	\alpha_{s,n_f} (r^2) =  \frac{4\pi}{\beta_{0, n_f} \ln \Big(\frac{4 C^2}{r^2 \Lambda_{n_f}^2}\Big) }\, ,
\end{equation}
where
\begin{equation}
\label{eq:beta}
	\beta_{0,n_f} = 11 - \frac{2}{3} n_f \, .
\end{equation}
Here, the constant $C^2$ under the logarithm accounts for the uncertainty inherent to the Fourier transform from momentum space, where the original calculation of the quark part of the $\beta$ function was performed \cite{Kovchegov:2006wf, Balitsky:2006wa}, to coordinate space. It will be one of the free parameters in the fits. 

In both our previous analysis \cite{Albacete:2009fh} and for the fits in subsection 4.1 only light quarks were taken as contributing to the DIS cross section. In this case, only fluctuations of the virtual photon wavefunction in Eq. (\ref{dm1}) into dipoles of light quark flavor were included in the calculation. Consistently, only light quark loops should be included in the calculation of the running coupling Eq. (\ref{eq:rc}). Thus, the number of active flavors in  Eq. (\ref{eq:rc}) is taken to be fixed and equal  to the number of light quarks  $n_f=3$. 

Since the rcBK equation is an integro-differential equation where the phase space for all dipole sizes is explored, including arbitrarily large dipole sizes (which correspond to emission of gluons with arbitrarily small transverse momenta), a prescription to regulate the coupling in the infrared is needed. We freeze the coupling to two constant values $\alpha_{fr}=0.7$ and 1 for dipole sizes larger than the scale at which the running coupling reaches $\alpha_{fr}$.

When heavy quark (charm and beauty) contributions are included in the calculation of the DIS cross section, as it is the case for the fits in subsection \ref{heavy}, fluctuations of the virtual photon wavefunction in Eq. (\ref{dm1}) into dipoles of heavy quark flavor are allowed. Accordingly, such contributions should be accounted for  in the computation of the running coupling  Eq. (\ref{eq:rc}). Thus, the number of active flavors $n_f$ in  Eq. (\ref{eq:rc}) should be set to the number of quark flavors lighter than the momentum scale associated with the scale $r^2$ at which the coupling is evaluated $\mu^2= 4 C^2/r^2$. The setup of this variable flavor scheme is completed by matching the branches of the coupling with adjacent $n_f$ at the scale corresponding to the quark masses 
$r_\star^2 = 4 C^2 / m_f^2$. For the 1-loop accuracy at which the coupling Eq. (\ref{eq:rc}) is evaluated, the matching condition is simply given by 
\begin{equation}
\label{eq:matchcond}
	\alpha_{s, n_f-1} (r_\star^2) = \alpha_{s, n_f} (r_\star^2)\, ,
\end{equation}
which results in 
\begin{equation}
\label{eq:lambdas}
	 \Lambda_{n_f -1} =(m_f)^{1-\frac{\beta_{0, n_f}}{\beta_{0,n_f-1}}}	 \, (\Lambda_{n_f})^{\frac{\beta_{0,n_f}}{\beta_{0, n_f-1}}}\, .
\end{equation}
The values of the $\Lambda_{n_f}$, $\Lambda_{3}$ in the fixed $n_f$ scheme and $\Lambda_{3}$, $\Lambda_{4}$, and $\Lambda_{5}$ for variable $n_f$ are determined by using an experimentally measured value of $\alpha_s$ as reference. It is a well known fact that the running of the QCD coupling evaluated to 1-loop is of insufficiently accuracy to describe the experimental observed coupling evolution. Thus, different choices of reference measurement will result in slightly different values for the  $\Lambda_{n_f}$. 
To take into account such uncertainty, in some of the fits we will use as reference point the experimentally measured value of $\alpha_s$ at the $Z^0$ mass, whereas in other fits the measured value of the coupling at the $\tau$ mass will be taken as the reference scale.

\subsection{Initial conditions for the evolution}
\label{ic}

Finally, to complete all the ingredients needed for the calculation of the reduced cross section \eq{rcs} we need to specify the initial conditions for the rcBK evolution equation \eq{bkrun}. Similarly to our previous work we consider GBW initial conditions, inspired in the phenomenological model of \cite{Golec-Biernat:1998js}:
\beq {\cal N}^{GBW}(r,x\!=\!x_0)=
1-\exp{\left[-\frac{\left(r^2\,Q_{s\,0}^2\right)^{\gamma\,}}{4}\right]}\,,
\label{gbw}
\eeq 
and MV initial conditions, which originate from a semiclassical calculation of multiple rescatterings \cite{McLerran:1997fk}:
 \beq {\cal
  N}^{MV}(r,x\!=\!x_0)=1-\exp{\left[-\frac{\left(r^2\,Q_{s\,0}^2\right)^{\gamma\,}}{4}
    \ln{\left(\frac{1}{r\,\Lambda}+e\right)}\right]}\, .
\label{mv}
\eeq
The physical meaning of the different parameters in Eqs (\ref{gbw}) and (\ref{mv}) is the following: $Q_{s0}$ is the saturation scale at the largest value of $x$ considered in the analysis, $x_0=0.01$, while $\gamma$ is an additional parameter that controls the steepness of the fall-off of the dipole amplitude with decreasing $r$.  It should be noted that the factor $\Lambda$ under the logarithm in the MV initial conditions corresponds to the infrared cutoff of the dipole-nucleon cross section at the level of two gluon exchange or in the semiclassical limit. Thus, it does not need to be equal to the $\Lambda_{n_f}$ in the running of the coupling. However, we opt to set it equal to $\Lambda_3$.

In order to further explore the space of initial conditions we shall consider a third family of i.c., the scaling i.c. which is generated by the evolution itself. It is a well known result that the asymptotic solutions of the rcBK equation are universal, i.e., they are independent of the initial conditions \cite{Iancu:2002tr,Albacete:2004gw,Mueller:2002zm, Albacete:2007yr}. Moreover, such asymptotic solutions present the feature of scaling, i.e. they do no longer depend on two kinematic variables $r$ and $Y$, but rather on a single dimensionless scale, the scaling variable $\tau=r\,Q_s(Y)$. In other words, the evolution generates a universal shape for the dipole amplitude at asymptotically large rapidities
\begin{equation}
\mathcal{N}(r,Y\gg1)\to \mathcal{N}^{scal}(\tau=r\,Q_s(Y)).
\label{scal}
\end{equation}
Since the analytic form of the universal shape $\mathcal{N}^{scal}$ is not known, the implementation of the scaling i.c. is done numerically: we solve \eq{bkrun} up to large rapidities, which we set to be $Y=80$. Then the obtained solution is rescaled by the corresponding value of the saturation scale, i.e we replace $\tau=r\,Q_s(Y)\to r\,Q_{s0}$ in \eq{scal}, where $Q_{s0}$ carries again the meaning of initial saturation scale at $x=x_0$ Thus, the scaling i.c. is essentially a one-parameter family of solutions, the only free parameter being the initial saturation scale.

\subsection{Parameters for fits with heavy quarks}
\label{hq}

As discussed at the beginning of this section, we replace the two-dimensional integral over impact parameter in \eq{dm1} by a dimensionful scale $\sigma_0$ which sets the normalization and can be interpreted as the average transverse size of the proton.  
However, it is not clear a priori whether such average area should be the same for quarks (valence or sea) and gluons. Indeed it has been suggested that the glue distribution inside nucleons may be located inside hot spots of small radius $\sim 0.2\div0.3$ fm \cite{Kopeliovich:1999am}. Also, data on the exponential slope of the momentum transfer dependence of exclusive vector meson production (see \cite{Marage:2009xz} and references therein) provide further support the picture of a smaller effective area for gluons than for valence quarks. Here we take as a working hypothesis the possibility that the effective transverse size of the heavy quark distribution, which we expect to follow the gluon one, may be different to that of light quarks. Accordingly, we introduce two different normalization constants for the total cross section, one for charm and beauty, $\sigma_0^{heavy}$  and other for the three light quarks, $\sigma_0$: 
 \begin{eqnarray}
  \sigma_{T,L}(x,Q^2)=\sigma_0\sum_{f=u,d,s}\int_0^1 dz \,d{\bf r}\,\vert
  \Psi_{T,L}^f(e_f,m_f,z,Q^2,{\bf r})\vert^2\,
  {\cal N}^{light}({\bf r},x)\,\nonumber \\
  +\,\sigma_0^{heavy}\sum_{f=c,b}\int_0^1 dz \,d{\bf r}\,\vert
  \Psi_{T,L}^f(e_f,m_f,z,Q^2,{\bf r})\vert^2\,
  {\cal N}^{heavy}({\bf r},x)\,.
\label{dm2}
\end{eqnarray}
As we shall discuss in section 4, such assumption is not only a physically well motivated one, but it turns out to be necessary in order to attain a good description of data, and also for the stability of the fits with respect to the inclusion or not of the heavy quark contribution. Finally, the superscripts $light$ and $heavy$ in the dipole scattering amplitudes in \eq{dm2} refers to the fact that we may consider different initial values of the parameters in the initial condition for light and heavy quarks. 
%This additional assumption would duplicate the number of free parameters but turns out not to be mandatory for a good description of data, as discussed in section 4.

\subsection{Summary of the theoretical setup and free parameters}
In summary, we will calculate the reduced cross section and the charm and beauty contribution to the inclusive structure functions according to the dipole model under the
translational invariant approximation \eq{dm2}.  The small-$x$
dependence is completely described by means of the BK equation
including running coupling corrections, Eqs. (\ref{bkrun}-\ref{kbal}),
for which three different initial conditions GBW, MV and scaling are considered. 
All in all, the free parameters to be fitted to experimental data are:
\begin{itemize}
\item $\sigma_0\,$: The total normalization of the cross section in \eq{dm2}.
\item $Q_{s\,0}^2\,$: The saturation scale of the proton at the highest
  experimental value of Bjorken-$x$ included in the fit, $x_0=10^{-2}$, in Eqs. (\ref{gbw}) and (\ref{mv}).
\item $C^2$: The parameter relating the running of the
  coupling in momentum space to the one in dipole size in \eq{eq:rc}.
\item $\gamma\,$: The anomalous dimension of the initial condition for
  the evolution in Eqs. (\ref{gbw}) and (\ref{mv}).
\end{itemize}
The fits with heavy quarks introduce additional free parameters, $\sigma_0^{heavy}$, $Q_0^{heavy}$ and $\gamma^{heavy}$, with physical meaning analogous to that of the corresponding parameters listed above.

\section{Numerical method and experimental data}
\label{numer}
The fit includes data on different observables and from different experiments:

$\bullet$ Data on the inclusive structure function $F_2$ by the E665 \cite{Adams:1996gu} (FNAL), the NMC \cite{Arneodo:1996qe} (CERN-SPS) collaborations.

$\bullet$ Data for the reduced cross section $\sigma_r$ from the combined analysis of the H1 and ZEUS collaborations \cite{:2009wt}. 

$\bullet$ Data on the charm contribution to the total structure function $F_{2c}$ \cite{Aktas:2004az, Aktas:2005iw, Breitweg:1997mj, Adloff:2001zj, Adloff:1996xq}. Even if the beauty contribution to the reduced cross sections is considered, we do not include in the fits the few available data on $F_{2b}$. They have large error bars and influence very little the fit output.

We have considered data for $x\leq10^{-2}$ and for all available values of $Q^2$, $0.045 \ {\rm GeV}^2\le Q^2 \le 50 \ {\rm GeV}^2$. 
The only published direct measurement of the longitudinal structure
function $F_L(x,Q^2)$ were obtained recently by the H1 and ZEUS collaborations
\cite{:2008tx,Chekanov:2009na}, and it is {\it not} included in the fit.
All in all, 325 data points are included in the fits with only light quarks and 329 data points in fits including heavy quarks and the results for $F_{2c}$. It should be noted that data points not fulfilling the condition $\tilde{x}<10^{-2}$ see, \eq{xtilde}, more constraining when charm quark is considered, were excluded from the data set, hence the relatively small difference in the total number of data points for fits with or without $F_{2c}$ data. 
Statistical and systematic uncertainties were added in quadrature, and
relative normalization uncertainties not considered. 
It was checked in \cite{:2009wt} that for the DGLAP analysis performed there, this rough treatment of uncertainties did not result in sizable differences with respect to the rigorous treatment of statistical, systematic and normalization uncertainties separately which is far more demanding on computer resources. Since the minimization algorithms
require a large number of calls to the function we have implemented a
parallelization of the numeric code. Finally, the BK
evolution equation including running coupling corrections is solved
using a Runge-Kutta method of second order with rapidity step $\Delta
h_y=0.05$.  For further details see \cite{Albacete:2007yr}.

\section{Results} \label{results} 
\subsection{Fits with only light quarks}
\TABLE{
\begin{tabular}{|l||c||c||c|c|c|c|c|}
\hline
&  fit   & $\frac{\chi^2}{d.o.f}$ & $Q_{s0}^2$ & $\sigma_0$ & $\gamma$ & $C$ & m$_l^2$\\
\hline
\hline
 & GBW   & & & & & & \\
\hline
a & $\alpha_{fr}=0.7$ &  1.226 & 0.241 & 32.357 & 0.971 & 2.46 & fixed\\
\hline
a' & $\alpha_{fr}=0.7$ ($\Lambda_{m_{\tau}}$)  & 1.235 & 0.240 & 32.569 & 0.959 & 2.507 & fixed\\
\hline
b & $\alpha_{fr}=0.7$  & 1.264 & 0.2633 & 30.325 & 0.968 & 2.246 & 1.74E-2\\
\hline
c & $\alpha_{fr}=1$   & 1.279 & 0.254 & 31.906 & 0.981 & 2.378 & fixed\\
\hline
c' & $\alpha_{fr}=1$ ($\Lambda_{m_{\tau}}$)  & 1.244 & 0.2329 & 33.608 & 0.9612 & 2.451 & fixed\\
\hline
d &  $\alpha_{fr}=1$   & 1.248 & 0.239 & 33.761 & 0.980 & 2.656 & 2.212E-2\\
\hline
%a & only comb hera &  1.332 & 0.241 & 32.712 & 0.948 & 2.58 & fixed\\
%\hline
\hline
& MV   & & & & & & \\
\hline
e & $\alpha_{fr}=0.7$  & 1.171 & 0.165 & 32.895 & 1.135 & 2.52 & fixed\\
\hline
f &  $\alpha_{fr}=0.7$  & 1.161 & 0.164 & 32.324 & 1.123 & 2.48 & 1.823E-2\\
\hline
g & $\alpha_{fr}=1$  & 1.140 & 0.1557 & 33.696 & 1.113 & 2.56 & fixed\\
\hline
h & $\alpha_{fr}=1$  & 1.117 & 0.1597 & 33.105 & 1.118 & 2.47 & 1.845E-2\\
\hline
h' & $\alpha_{fr}=1$  ($\Lambda_{m_{\tau}}$)  & 1.104 & 0.168 & 30.265 & 1.119 & 1.715 & 1.463E-2 \\
\hline
\end{tabular}

\label{tablight}
\caption{Parameters from fits with only light quarks to data with $x\leq10^{-2}$ and for all
  available values of $Q^2 \le 50 \
  {\rm GeV}^2$ for different initial conditions, fixed values of the coupling in the infrared $\alpha_{fr}=0.7$ and 1 and light quark masses either taken fixed $m_{l}=0.14$ GeV or left as a free parameter. Fits a', c' and h' correspond to taking the $\tau$ mass as reference scale for the running of the coupling. Units: $Q_{s0}^2$ and $m_l^2$ are in GeV$^2$ and $\sigma_0$ in mb.}
%\end{center}
%\end{table}
}
We first perform fits to data accounting only for the light quark contribution in \eq{dm1}, $f=u,d,s$. For these fits we take $n_f\!=\!3$ in the running of the coupling. As the default setting we adjust $\Lambda_{QCD}$ to reproduce  $\alpha_s(m_Z)$ correctly. We assessed the effect of this choice by also performing fits where $\alpha_s (m_{\tau})$ was alternatively taken as the reference, obtaining similarly good fits.
In all these fits we impose an upper cut $Q^2=50$ in the data. However, we note that the fits are stable after extending our analyses to data with larger values of $Q^2\le 100$ GeV$^2$, albeit at the prize of getting slightly worse $\chi^2/d.o.f.$.
The values of the free parameters obtained from the fits to data are presented in Table \ref{tablight}, and  a partial comparison with the experimental data for the reduced cross section is presented in Fig. 1 (left plot). Note that, with several combinations resulting in fits of very similar quality, both in this subsection and in the following one we will only show in the plots the results from some selected fits, not a full survey of them. Several comments are in order:

First, all the different fits with MV or GBW initial conditions yield a good $\chi^2/d.o.f\le 1.28$, with a best fit $\chi^2/d.o.f\!=\!1.104$, labeled h' in table \ref{tablight}, obtained with MV initial condition, $\alpha_{fr} = 1$, andÊ $\alpha_s (m_{\tau})$ as the reference value for the running coupling. 
The quality of the fits is remarkably good provided the tiny error bars in the new data on reduced cross sections (error bars are in most cases smaller than the symbols used in the plot). In turn, it was not possible to find any good fit to data using the {\it scaling} initial condition \eq{scal}. Most likely, this is due to the much faster evolution speed featured by the {\it scaling} initial conditions, compared to the GBW or MV ones, for which pre-asymptotic effects slow down the evolution considerably. Moreover, the MV i.c. tend to systematically yield better fits than the GBW one. This can be taken as an indication that the semiclassical resummation of multiple scattering underlying the MV formula is indeed a good estimate of the initial condition. 
 
 Next, the sensitivity of the fits to non-perturbative aspects of our calculation encoded in the parameters $\alpha_{fr}$, $C$, the reference scale to determine $\Lambda_{n_f}$ or the light quark masses $m_l$ (which acts as an effective IR cutoff for the $\Psi^{\gamma^*\to q\bar{q}}$ wavefunction) is rather small, as shown by the little variation of the fit parameters under changes in the latter and on whether they are left as free fit parameters or not. In particular, the value at which the coupling is regularized in the infrared, either 1 or 0.7 does not affect much the fit output. Also, when the light quark mass is left as a free parameter it tends to acquire a final value very close to the default one, $m_l=140$ MeV. This gives us confidence that the good agreement with data is indeed driven by the small-$x$ dynamics encoded in the rcBK equation rather than being due to a fine tuning of the remaining parameters.
Importantly, the fits parameters are similar, in all cases, to those obtained in our previous work \cite{Albacete:2009fh}.

%\begin{table}[h]
%\begin{center}
%%%%%%%%%%%%%%%%%%%%%%%%%%%%%
%\FIGURE{
\hspace{-2cm}
\begin{figure}[ht]
\begin{center}
 \includegraphics[height=7.7cm]{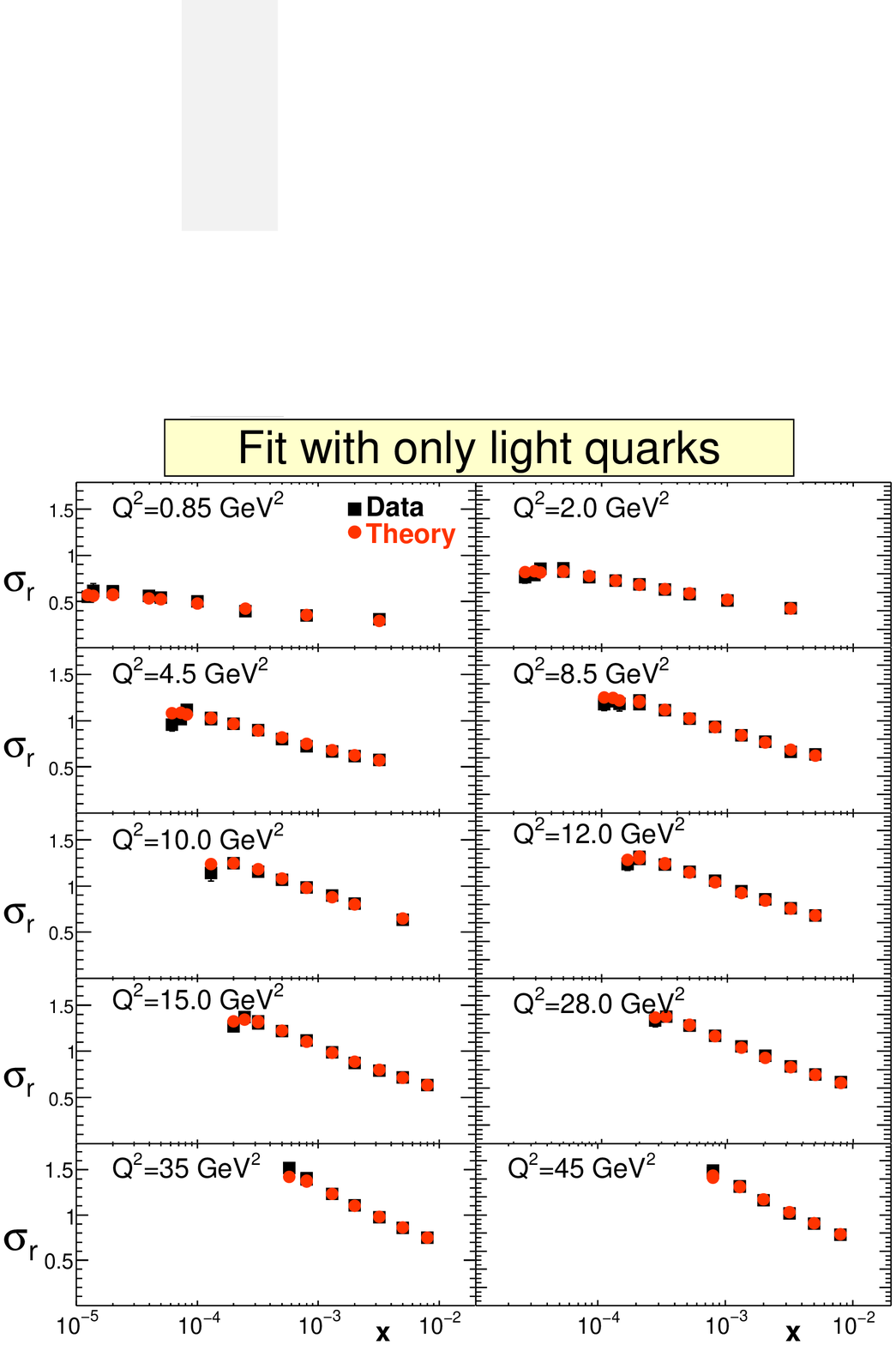}%c
\includegraphics[height=7.7cm]{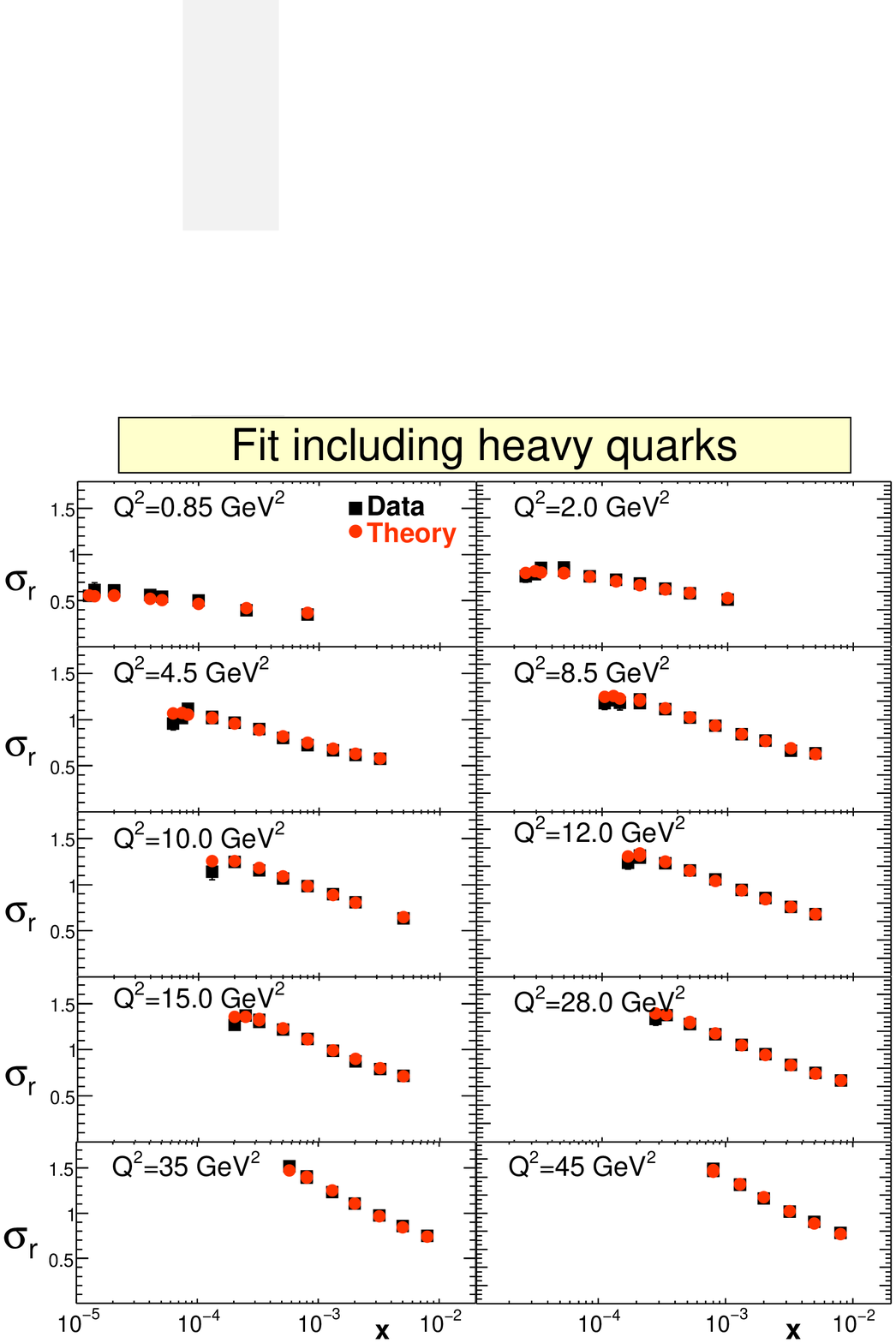}
\end{center}
\caption{Comparison of experimental data for the reduced cross sections (black squares) in different $Q^2$ bins with our results (red circles). The results in the left plot correspond to a fit with only light flavors and GBW initial condition, entry (a) in Table 1. The results in the right plot include the contribution of charm and beauty quarks and correspond to fit (a') in Table 2. }
\label{sr_light}
\end{figure}

\subsection{Inclusion of heavy quarks in the fits.}
\label{heavy}
\TABLE{
\begin{tabular}{|l||l||c||c|c|c|c|c|c|c|c|}
\hline
%$Q^2_{cut}=50$GeV$^2$&  & & & & & & & & &\\
%\hline
%\hline
& fit  & $\frac{\chi^2}{d.o.f}$ & $Q_{s0}^2$ & $\sigma_0$ & $\gamma$ & $Q_{s0c}^2$ & $\sigma_{0c}$ & $\gamma_c$& $C$ & $m_{l}^2$ \\
\hline
\hline
 & GBW   & & & & & & & &  & \\
\hline
a & $\alpha_{fr}\!=\!0.7$  &  1.269 & 0.2294 & 36.953 & 1.259 & 0.2289 & 18.962 & 0.881 & 4.363 & fixed \\
\hline
a' & $\alpha_{fr}\!=\!0.7$ ($\Lambda_{m_{\tau}}$) & 1.302 & 0.2341 & 36.362 & 1.241 & 0.2249 & 20.380 & 0.919 & 7.858 & fixed \\
\hline
b & $\alpha_{fr}\!=\!0.7$ & 1.231 & 0.2386 & 35.465 & 1.263 & 0.2329 & 18.430 & 0.883 & 3.902 & 1.458E-2 \\
\hline
c & $\alpha_{fr}\!=\!1$ & 1.356 & 0.2373 & 35.861 & 1.270 & 0.2360 & 13.717 & 0.789 & 2.442 & fixed \\
\hline
d & $\alpha_{fr}\!=\!1$ & 1.221 & 0.2295 & 35.037 & 1.195 & 0.2274 & 20.262 & 0.924 & 3.725 & 1.351E-2 \\
\hline
\hline
& MV   & & & & & & & & & \\
\hline
e & $\alpha_{fr}\!=\!0.7$   & 1.395 & 0.1673 & 36.032 & 1.355 & 0.1650 & 18.740 & 1.099 & 3.813 & fixed \\
\hline
f & $\alpha_{fr}\!=\!0.7$ & 1.244 & 0.1687 & 35.449 & 1.369 & 0.1417 & 19.066 & 1.035 & 4.079 & 1.445E-2 \\
\hline
g & $\alpha_{fr}\!=\!1$ & 1.325 & 0.1481 & 40.216 & 1.362 & 0.1378 & 13.577 & 0.914 & 4.850 & fixed \\
\hline
h & $\alpha_{fr}\!=\!1$ & 1.298 & 0.156 & 37.003 & 1.319 &  0.147 & 19.774 & 1.074 & 4.355 & 1.692E-2 \\
\hline
\end{tabular}

\label{tabheavy}
\caption{Parameters from fits including charm and beauty contributions to data with $x\leq10^{-2}$ and and $Q^2 \le 50 \
  {\rm GeV}^2$ for different initial conditions and fixed values of the coupling in the infrared $\alpha_{fr}=0.7$ and 0.1. Light quark masses are fixed to $m_{l}=0.14$ GeV in some fits and left as a free parameter in others. The fit a' corresponds to taking the $\tau$ mass as reference scale for the running of the coupling. The units: $Q_{s0(c)}^2$ and $m_l^2$ are GeV$^2$, while those of $\sigma_{0(c)}$ are mb.}
}

%%%%%%%%%%%%%%%%%%%%%%%%%%%%%
%\FIGURE{
\begin{figure}[ht]
\begin{center}
\includegraphics[height=10cm]{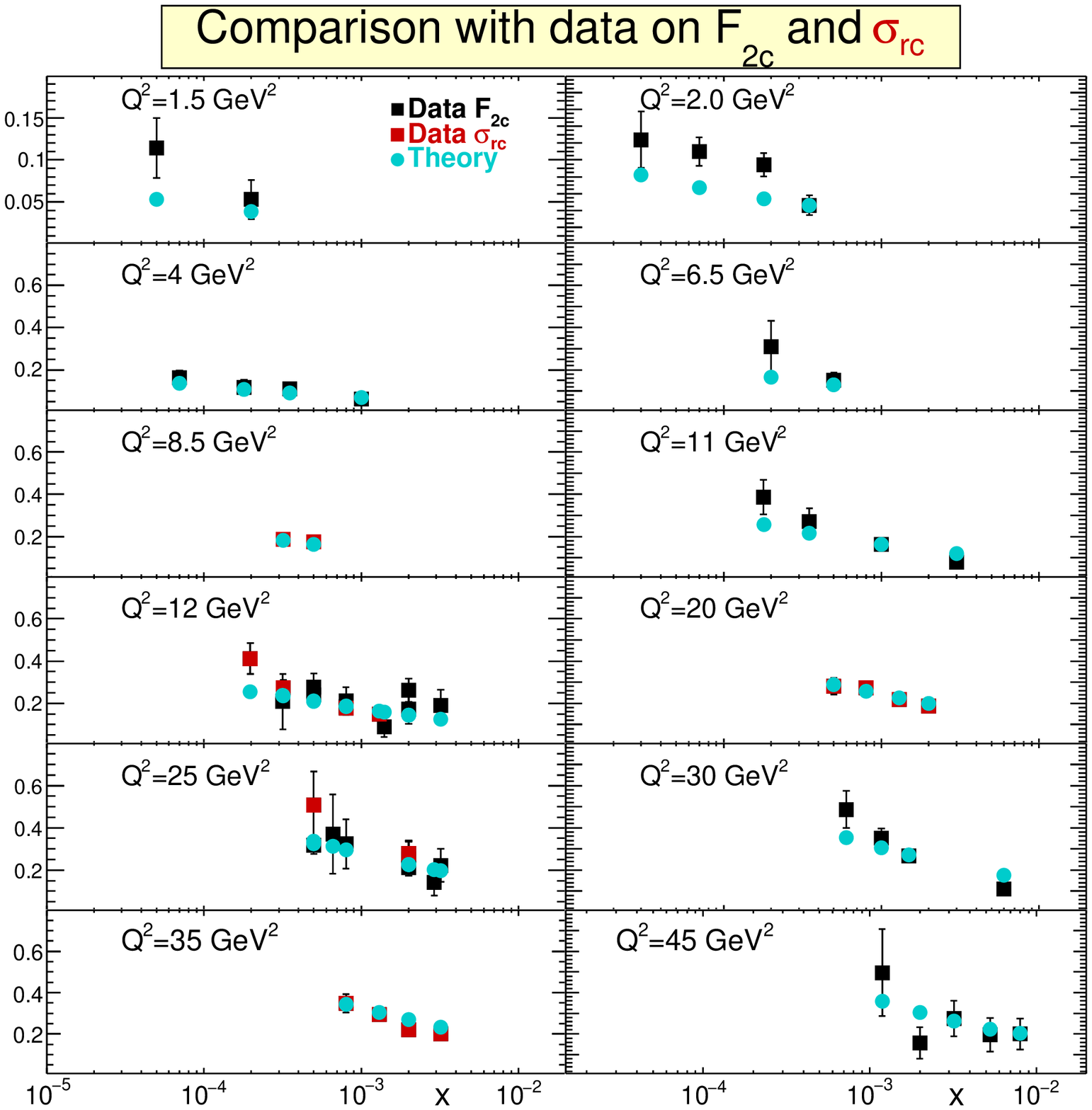}%c
\end{center}
\caption{Comparison of experimental data for $F_{2c}$ (black squares) and $\sigma_{rc}$ (red squares)  in different $Q^2$ bins with our results (cyan circles), corresponding to fit (a') in Table 2.}
\label{f2c}
\end{figure}
%%%%%%%%%%%%%%%%%%%%%%%%%%%%%

In this section we present the fits performed including the contribution of charm and beauty quarks into \eq{dm1}. As discussed earlier, we find that in order to obtain a good description of data while keeping the stability of the fit parameters for light quarks it is necessary to assume that the overall normalization of the heavy quark contribution to the reduced cross section is different to the one for light quarks. This translates into the introduction of a new free parameter, $\sigma_{0}^{heavy}$, which turns out to be smaller than $\sigma_0$, the corresponding normalization for light quarks. This can be interpreted as the average radius of the heavy quark distribution being smaller than the one for light quarks. In principle, there is no reason a priori why such average radius should be the same for charm and beauty quarks. On the contrary, one may expect a smaller size of the effective beauty distribution on account of its larger mass. This would suggest the introduction of two different normalization parameters for charm and beauty $\sigma_{0c}$ and $\sigma_{0b}$, as well as, maybe, different initial conditions for the evolution for each heavy quark flavor. However, the paucity of data on $F_{2b}$ or related observables able to independently constrain the free parameters associated to the beauty quark prevents us of carrying out a more detailed characterization of its contribution to the data included in the fit. Thus, we assume that the free parameters associated to heavy quarks, including the overall normalization, is the same for charm and beauty. We have checked that such assumption has a very little effect on the fit output by completely removing the beauty contribution to $F_2$ and $\sigma_r$. However, we finally decided to include it in the fits in order to be consistent with the variable flavor scheme used for the running of the coupling, which allows the contribution of dynamical $b$ quarks to the QCD beta function.

Our fit results are shown in Table \ref{tabheavy}, and a comparison with data for $\sigma_{r}$ is shown in the right plot Fig 1. We obtain an equally good description of data as with fits with only light quarks, as can be seen comparing the left and right plots in Fig 1. However, the $\chi^2/d.o.f. \lesssim 1.4$ are slightly larger than for the fits with only light quarks.  This is maybe due to what seems to be a systematic deviation between different data sets on $F_{2c}$ and the charm contribution to the reduced cross section $\sigma_{rc}$, as can be observed in Fig 2, where we compare our results with experimental data.
The arguments presented before on the stability of the fits with respect to variations in the infrared regulation of the coupling or the reference scale to determine $\Lambda_{QCD}$ also hold in the case of fits with heavy quarks.  On the other hand, when left as a free parameter the mass of the light quark tends to acquire a smaller value than it did in the fits with only light quarks. Concerning the initial conditions for the evolution, they are very similar for light and heavy quarks. In particular, the corresponding initial saturation scales, $Q_{s0}$ and $Q_{s0c}$ take on very similar values in all fits. However, the steepness of the initial condition encoded in the parameter $\gamma_{(c)}$ is systematically larger for light than for heavy quarks for both GBW and MV initial conditions.

\subsection{Comparison with $F_L$}
%%%%%%%%%%%%%%%%%%%%%%%%%%%%%
%\FIGURE{
\begin{figure}[ht]
\begin{center}
\includegraphics[height=9cm]{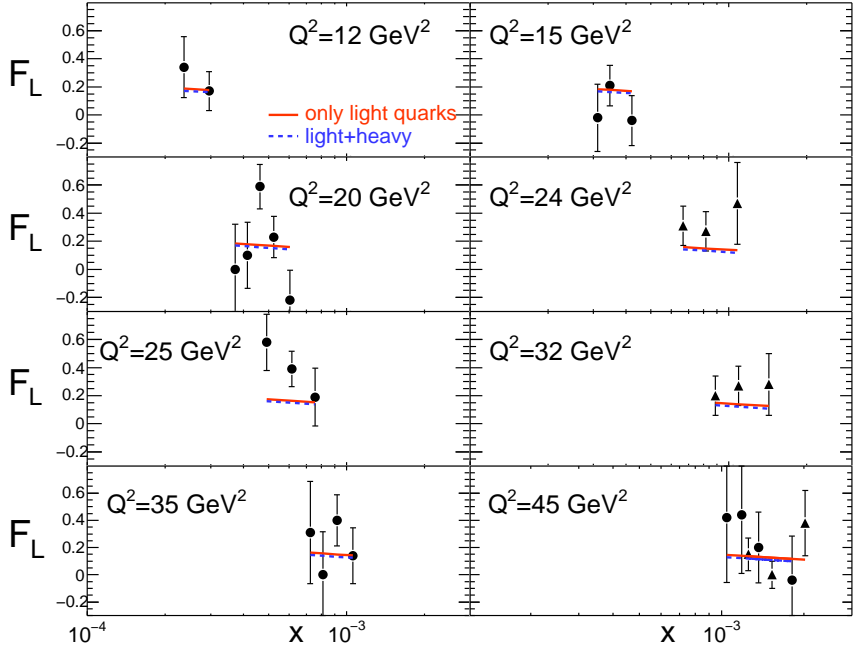}%c
\end{center}

\caption{Comparison of experimental data for $F_{L}$ from the H1 (full circles) and ZEUS (triangles) collaborations with the theoretical results corresponding to a fit with only light quarks and MV i.c.  (solid line, labeled (e) in Table 1) and a fit with MV i.c. and including heavy quarks (dashed line, labeled (a) in Table 2)  . }
\label{sr_light}
\end{figure}

In Fig 3. we present a comparison of our results for the longitudinal structure function $F_L$ with the available data at small-$x$ and for different $Q^2$ bins. The theoretical results were obtained using the dipole parametrizations corresponding to fits (e) and (a) in Tables 1 and 2 respectively, although we have checked that all the others provide equally good comparisons with data. The agreement with data is good, provided the relatively large error bars in experimental data.

%%%%%%%%%%%%%%%%%%%%%%%%%%%%%
 
\section{Conclusions} 
\label{conclusions}
In this paper we have presented an analysis of the available data on the several inclusive structure functions and reduced cross section measured in e+p collisions at small-$x$. This proves the ability of the rcBK equation, the main dynamical ingredient in our approach, to account for the $x$-dependence of the available data, including the high quality data on reduced cross sections provided by the combined analysis of the H1 and ZEUS Collaborations. We thus offer additional indications for the presence of non-linear {\it saturation} effects in present data and, thereby, sharpen the CGC approach to high-energy QCD scattering as a practical phenomenological tool.
We have also shown how the inclusion of heavy quarks, both at the level of their contribution of the QCD beta function in the running of the coupling as well as to the total $\gamma^*$-proton cross sections can be  naturally incorporated in the dipole formalism under the assumption of a smaller size of the heavy quark effective distribution. The dipole scattering amplitude solving the rcBK equation stemming from the parameter sets in Tables 1 and 2 shall be publicly available in the form of numeric Fortran routines at the website http://www-fp.usc.es/phenom/software.html.

%%%%%%%%%%%%%%%%%%%%%%%%%%%%%%%%%%%%%%%%%%%%%%%%%%%%%%%%%%%%%%%%%%%%%%%%%%%%%%%

\section*{Acknowledgments} 

We thank F. Gelis and G. Soyez for helpful discussions. The work of JLA is supported by a Marie Curie Intra-European Fellowship (FP7- PEOPLE-IEF-2008), contract No. 236376.  JLA also acknowledges support from the CAPES-COFECUB program and the warm hospitality by the Departamento de F\'{i}sica Te\'orica of Universidade Federal do Rio de Janeiro, where part of this project was carried out. 
This work is supported by the Ministerio de Ciencia e Innovaci\'on (proyectos FPA2008-01177 y FPA2009-06867-E), Xunta de Galicia (Conseller\'{\i}a de Educaci\'on and projects PGIDIT10PXIB \-206017PR (NA) and INCITE08PXINCITE08PXIB296116PR (CAS)), y proyecto Consolider-Ingenio 2010 CPAN (CSD2007-00042) and by the European Commission grant
PERG02-GA-2007-224770 and by Funda\c c\~ao para a Ci\^encia e a Tecnologia
of Portugal under project CERN/FP/109356/2009 and contract CIENCIA 2007
(JGM). CAS is a Ram\'on y Cajal researcher. The work of PQA is funded by the French ANR under contract ANR-09-BLAN-0060ANR-09-BLAN-0060.

%This work has been supported by Ministerio de Ciencia e Innovaci\'on of Spain under
%projects FPA2005-01963, FPA2008-01177 and contracts Ram\'on y Cajal
%(CAS); by Xunta de Galicia (Conseller\'{\i}a de Educaci\'on and
%Conseller\'\i a de Innovaci\'on e Industria -- Programa Incite) (NA
%and CAS); by the Spanish Consolider-Ingenio 2010 Programme CPAN
%(CSD2007-00042) (NA and CAS); by the European Commission grant
%PERG02-GA-2007-224770 (CAS);

%%%%%%%%%%%%%%%%%%%%%%%%%%%%%%%%%%%%%%%%%%%%%%%%%%%%%%%%%%%%%%%%%%%%%%%%%%%%%%

%\bibliography{ref2}{}

\begin{thebibliography}{10}

\bibitem{Jalilian-Marian:1997gr}
J.~Jalilian-Marian, A.~Kovner, A.~Leonidov, and H.~Weigert,
\newblock Phys. Rev. {\bf D59}, 014014 (1998), hep-ph/9706377.

\bibitem{Jalilian-Marian:1997dw}
J.~Jalilian-Marian, A.~Kovner, and H.~Weigert,
\newblock Phys. Rev. {\bf D59}, 014015 (1998), hep-ph/9709432.

\bibitem{Kovner:2000pt}
A.~Kovner, J.~G. Milhano, and H.~Weigert,
\newblock Phys. Rev. {\bf D62}, 114005 (2000), hep-ph/0004014.

\bibitem{Weigert:2000gi}
H.~Weigert,
\newblock Nucl. Phys. {\bf A703}, 823 (2002), hep-ph/0004044.

\bibitem{Iancu:2000hn}
E.~Iancu, A.~Leonidov, and L.~D. McLerran,
\newblock Nucl. Phys. {\bf A692}, 583 (2001), hep-ph/0011241.

\bibitem{Ferreiro:2001qy}
E.~Ferreiro, E.~Iancu, A.~Leonidov, and L.~McLerran,
\newblock Nucl. Phys. {\bf A703}, 489 (2002), hep-ph/0109115.

\bibitem{Balitsky:1996ub}
I.~Balitsky,
\newblock Nucl. Phys. {\bf B463}, 99 (1996), hep-ph/9509348.

\bibitem{Kovchegov:1999yj}
Y.~V. Kovchegov,
\newblock Phys. Rev. {\bf D60}, 034008 (1999), hep-ph/9901281.

\bibitem{Nikolaev:1990ja}
N.~N. Nikolaev and B.~G. Zakharov,
\newblock Z. Phys. {\bf C49}, 607 (1991).

\bibitem{Mueller:1989st}
A.~H. Mueller,
\newblock Nucl. Phys. {\bf B335}, 115 (1990).

\bibitem{Golec-Biernat:1998js}
K.~Golec-Biernat and M.~{W{\"u}sthoff},
\newblock Phys. Rev. {\bf D59}, 014017 (1998), hep-ph/9807513.

\bibitem{Iancu:2003ge}
E.~Iancu, K.~Itakura, and S.~Munier,
\newblock Phys. Lett. {\bf B590}, 199 (2004), hep-ph/0310338.

\bibitem{Gotsman:2002yy}
E.~Gotsman, E.~Levin, M.~Lublinsky, and U.~Maor,
\newblock Eur. Phys. J. {\bf C27}, 411 (2003), hep-ph/0209074.

\bibitem{Albacete:2005ef}
J.~L. Albacete, N.~Armesto, J.~G. Milhano, C.~A. Salgado, and U.~A. Wiedemann,
\newblock Eur. Phys. J. {\bf C43}, 353 (2005), hep-ph/0502167.

\bibitem{Kowalski:2003hm}
H.~Kowalski and D.~Teaney,
\newblock Phys. Rev. {\bf D68}, 114005 (2003), hep-ph/0304189.

\bibitem{Kowalski:2006hc}
H.~Kowalski, L.~Motyka, and G.~Watt,
\newblock Phys. Rev. {\bf D74}, 074016 (2006), hep-ph/0606272.

\bibitem{Goncalves:2006yt}
V.~P. Goncalves, M.~S. Kugeratski, M.~V.~T. Machado, and F.~S. Navarra,
\newblock Phys. Lett. {\bf B643}, 273 (2006), hep-ph/0608063.

\bibitem{Balitsky:2006wa}
I.~I. Balitsky,
\newblock Phys. Rev. D {\bf 75}, 014001 (2007), hep-ph/0609105.

\bibitem{Kovchegov:2006vj}
Y.~Kovchegov and H.~Weigert,
\newblock Nucl. Phys. {\bf A} {\bf 784}, 188 (2007), hep-ph/0609090.

\bibitem{Gardi:2006rp}
E.~Gardi, J.~Kuokkanen, K.~Rummukainen, and H.~Weigert,
\newblock Nucl. Phys. {\bf A784}, 282 (2007), hep-ph/0609087.

\bibitem{Balitsky:2008zza}
I.~Balitsky and G.~A. Chirilli,
\newblock Phys. Rev. {\bf D77}, 014019 (2008), 0710.4330.

\bibitem{Albacete:2007yr}
J.~L. Albacete and Y.~V. Kovchegov,
\newblock Phys. Rev. {\bf D75}, 125021 (2007), arXiv:0704.0612 [hep-ph].

\bibitem{Albacete:2007sm}
J.~L. Albacete,
\newblock Phys. Rev. Lett. {\bf 99}, 262301 (2007), 0707.2545.

\bibitem{Albacete:2009ps}
J.~L. Albacete, N.~Armesto, J.~G. Milhano, and C.~A. Salgado,
\newblock (2009), 0906.2721.

\bibitem{Albacete:2009fh}
J.~L. Albacete, N.~Armesto, J.~G. Milhano, and C.~A. Salgado,
\newblock Phys. Rev. {\bf D80}, 034031 (2009), 0902.1112.

\bibitem{Betemps:2009ie}
M.~A. Betemps, V.~P. Goncalves, and J.~T. de~Santana~Amaral,
\newblock (2009), 0907.3416.

\bibitem{Albacete:2010ad}
J.~L. ALbacete and A.~Dumitru,
\newblock (2010), 1011.5161.

\bibitem{Albacete:2010bs}
J.~L. Albacete and C.~Marquet,
\newblock Phys. Lett. {\bf B687}, 174 (2010), 1001.1378.

\bibitem{Albacete:2010pg}
J.~L. Albacete and C.~Marquet,
\newblock Phys. Rev. Lett. {\bf 105}, 162301 (2010), 1005.4065.

\bibitem{Dittmar:2009ii}
M.~Dittmar {\em et~al.},
\newblock (2009), 0901.2504.

\bibitem{Caola:2009iy}
F.~Caola, S.~Forte, and J.~Rojo,
\newblock Phys. Lett. {\bf B686}, 127 (2010), 0910.3143.

\bibitem{Altarelli:1999vw}
G.~Altarelli, R.~D. Ball, and S.~Forte,
\newblock Nucl. Phys. {\bf B575}, 313 (2000), hep-ph/9911273.

\bibitem{Ciafaloni:2003rd}
M.~Ciafaloni, D.~Colferai, G.~P. Salam, and A.~M. Stasto,
\newblock Phys. Rev. {\bf D68}, 114003 (2003), hep-ph/0307188.

\bibitem{Capella:2000pe}
A.~Capella, E.~G. Ferreiro, C.~A. Salgado, and A.~B. Kaidalov,
\newblock Nucl. Phys. {\bf B593}, 336 (2001), hep-ph/0005049.

\bibitem{Armesto:2010ee}
N.~Armesto, A.~B. Kaidalov, C.~A. Salgado, and K.~Tywoniuk,
\newblock Phys. Rev. {\bf D81}, 074002 (2010), 1001.3021.

\bibitem{lhec2}
M.~K. {\em et.~al.},
\newblock EPAC'08, 11th European Particle Accelerator Conference  (2008).

\bibitem{EICwhite}
{{\em The Electron Ion Collider: A white paper}, BNL Report
  BNL-68933-02/07-REV, Eds. A. Deshpande, R. Milner and R. Venugopalan. }.

\bibitem{:2009wt}
H1, F.~D. Aaron {\em et~al.},
\newblock JHEP {\bf 01}, 109 (2010), 0911.0884.

\bibitem{:2009ut}
H1, F.~D. Aaron {\em et~al.},
\newblock Eur. Phys. J. {\bf C65}, 89 (2010), 0907.2643.

\bibitem{Adloff:1996xq}
H1, C.~Adloff {\em et~al.},
\newblock Z. Phys. {\bf C72}, 593 (1996), hep-ex/9607012.

\bibitem{Adloff:2001zj}
H1, C.~Adloff {\em et~al.},
\newblock Phys. Lett. {\bf B528}, 199 (2002), hep-ex/0108039.

\bibitem{Aktas:2005iw}
H1, A.~Aktas {\em et~al.},
\newblock Eur. Phys. J. {\bf C45}, 23 (2006), hep-ex/0507081.

\bibitem{Aktas:2004az}
H1, A.~Aktas {\em et~al.},
\newblock Eur. Phys. J. {\bf C40}, 349 (2005), hep-ex/0411046.

\bibitem{Breitweg:1997mj}
ZEUS, J.~Breitweg {\em et~al.},
\newblock Phys. Lett. {\bf B407}, 402 (1997), hep-ex/9706009.

\bibitem{Soyez:2007kg}
G.~Soyez,
\newblock Phys. Lett. {\bf B655}, 32 (2007), 0705.3672.

\bibitem{Berger:2010sh}
J.~Berger and A.~Stasto,
\newblock (2010), 1010.0671.

\bibitem{Kormilitzin:2010at}
A.~Kormilitzin and E.~Levin,
\newblock (2010), 1009.1468.

\bibitem{Nakamura:2010zzi}
Particle Data Group, K.~Nakamura {\em et~al.},
\newblock J. Phys. {\bf G37}, 075021 (2010).

\bibitem{Kovchegov:2006wf}
Y.~V. Kovchegov and H.~Weigert,
\newblock Nucl. Phys. {\bf A789}, 260 (2007), hep-ph/0612071.

\bibitem{McLerran:1997fk}
L.~D. McLerran and R.~Venugopalan,
\newblock Phys. Lett. {\bf B424}, 15 (1998), nucl-th/9705055.

\bibitem{Iancu:2002tr}
E.~Iancu, K.~Itakura, and L.~McLerran,
\newblock Nucl. Phys. {\bf A708}, 327 (2002), hep-ph/0203137.

\bibitem{Albacete:2004gw}
J.~L. Albacete, N.~Armesto, J.~G. Milhano, C.~A. Salgado, and U.~A. Wiedemann,
\newblock Phys. Rev. {\bf D71}, 014003 (2005), hep-ph/0408216.

\bibitem{Mueller:2002zm}
A.~H. Mueller and D.~N. Triantafyllopoulos,
\newblock Nucl. Phys. {\bf B640}, 331 (2002), hep-ph/0205167.

\bibitem{Kopeliovich:1999am}
B.~Z. Kopeliovich, A.~Schafer, and A.~V. Tarasov,
\newblock Phys. Rev. {\bf D62}, 054022 (2000), hep-ph/9908245.

\bibitem{Marage:2009xz}
H1 and ZEUS, P.~Marage,
\newblock (2009), 0911.5140.

\bibitem{Adams:1996gu}
E665, M.~R. Adams {\em et~al.},
\newblock Phys. Rev. {\bf D54}, 3006 (1996).

\bibitem{Arneodo:1996qe}
New Muon, M.~Arneodo {\em et~al.},
\newblock Nucl. Phys. {\bf B483}, 3 (1997), hep-ph/9610231.

\bibitem{:2008tx}
H1, F.~D. Aaron {\em et~al.},
\newblock Phys. Lett. {\bf B665}, 139 (2008), 0805.2809.

\bibitem{Chekanov:2009na}
ZEUS, S.~Chekanov {\em et~al.},
\newblock Phys. Lett. {\bf B682}, 8 (2009), 0904.1092.

\end{thebibliography}
%\bibliographystyle{JHEP}
%\bibliographystyle{h-physrev}

\end{document}